\documentclass{article}     
\usepackage{epsf}

\title{Transport and Elastic Properties of Fractal Media}

\author{Anthony P. Roberts \\ Faculty of Environmental Sciences, \\
Griffith University, Nathan, Qld.\ 4111, Australia \\ \\
Mark A. Knackstedt \\ Department of Applied Mathematics, \\
Research School of Physical Sciences and Engineering, \\
Australian National University,
Canberra ACT 0200, Australia}
\date{October 1995}
\begin{document}
\renewcommand\floatpagefraction{0.1}
\maketitle
\begin{picture}(0,0)
\put (57,216){\parbox{13.0cm}{\small \raggedleft %
{\em Appeared in} Physica A, Vol.\ 233 (1996) pages 848--858}}
\end{picture}

\begin{abstract}
We investigate the influence of fractal structure on material properties.
We calculate the statistical correlation functions of fractal media defined
by level-cut Gaussian random fields. This allows the modeling of both
surface fractal and mass fractal materials. Variational bounds on the
conductivity, diffusivity and elastic moduli of the materials are evaluated.
We find that a fractally rough interface has a relatively strong
influence on the properties of composites. In contrast a fractal
volume (mass) has little effect on material properties.
\end{abstract}

\section{Introduction}
In the past decade a number of important natural and manufactured
materials have been shown to exhibit microstructures with fractal
characteristics.  For example, it has been 
shown that sedimentary rocks have internal surfaces that exhibit
fractal roughness and in some cases the fractal regime may include the
entire pore volume~\cite{Wong86}.  This geometric complexity has a number of 
important implications for transport properties in porous rocks.
In another example it was shown that fractured surfaces of metals are fractal
in nature and proposed that the fractal dimensions of the surfaces 
could be correlated to the impact energy of the sample~\cite{Su95}.  
Finally low-weight polymeric gel materials (aerogels) give 
a unique combination of properties;
low-density and insulative, yet mechanically strong~\cite{Smirnov90}.  
Aerogels are good examples of 
fractal materials, made up of clusters generated from the aggregation of 
primary particles~\cite{Schaefer84}.  

While the concept of a fractal geometry provides a
description of disordered material structure, it is not clear if the
properties of these
materials are affected by fractal characteristics nor 
how one may design and optimise fractal structure for a given application. 
Microstructural characteristics other than fractal dimension 
(e.g., connectivity, coordination) may be more important when 
deriving Structure-Property relationships.
For example, in aerogels, the macroscopic connectivity
at low density is the key microstructural characteristic associated with
the mechanical strength of the material.  

To date there has been no explicit investigation of the influence of
fractal structure on material properties.
Theoretical studies~\cite{Kantor85,Adler93a,Schwartz89,Smirnov90,Sahimi93}
have relied
on simplistic models of fractal microstructure.  In general
the relationship of such models to real materials is unclear, and the
predictive ability of the models is qualitative in nature. In this paper we
investigate the influence of fractal structure on material properties.
We find that a fractally rough interface has a strong influence on 
material properties.  A fractal volume, in contrast, has little effect
on properties.

\section{Evaluation of Bounds}

For general random composites there is no exact method of predicting
properties. However a number of rigorous bounds, which depend on the
microstructure of the composites, have been derived. 
These include bounds on the conductivity~\cite{Beran65a,Milton81b}
and the bulk~\cite{Beran65b} and shear~\cite{McCoy,Milton82b} moduli
of composite materials (reviewed in Ref.~\cite{TorqRev91}).
The bounds are expressed~\cite{Milton81a} in terms of the
volume fractions and properties of each of the phases and two
microstructure parameters:
\begin{eqnarray}
\label{zeta}
\zeta_1&=&
\frac9{2pq}\int_0^\infty\frac{dr}{r} \int_0^\infty
\frac{ds}{s} \int_{-1}^1 du  P_2(u) 
\left( p_3(r,s,t)-\frac{p_2(r)p_2(s)}p \right) \\
\eta_1&=&\frac {5\zeta_1}{21}+
\frac{150}{7pq}\int_0^\infty\frac{dr}{r}
\int_0^\infty\frac{ds}{s}
\int_{-1}^1  du  P_4(u) \left( p_3(r,s,t)-\frac{p_2(r)p_2(s)}p \right)
\label{eta}
\end{eqnarray}
where $p_3(r,s,t)$ is the three point correlation function~\cite{Brown55},
$t^2=r^2+s^2-2rs u$ and $P_n(u)$ denotes the Legendre polynomial of order $n$.
Until recently
the use of the bounds has been restricted to `particulate' media, such
as uncorrelated overlapping spheres~\cite{TorqRev91}.
Such models have limited utility
in describing
the microstructure of composite materials.
We have shown that microstructure generated from level cuts of 
a random standing wave mimics the microstructure of a wide range
of real composite materials including 
polymer blends~\cite{Knackstedt95a}, porous rocks~\cite{Roberts95d}
and foamed solids~\cite{Roberts95b}.
We have subsequently 
evaluated the bounds for a variety of level-cut microstructures and
shown that it is possible to correlate quantitatively 
the effective physical properties of materials 
to their microstructure~\cite{Knackstedt95a,Roberts95b}.
Berk~\cite{Berk91} has shown how the level-cut scheme may be used to
model fractal interfaces, and the extension to volume fractals is simple.
Here we investigate the properties of these surface- and
volume-fractal materials using rigorous bounds.  This allows us to 
investigate the influence of fractal structure on material properties.

A key quantity in the characterization of two phase materials is
the two-point correlation function, $p_2(r)=p(1-p)\gamma(r)+p^2$, where
$p$ is the volume fraction of the first phase
and $\gamma(r)$ is the normalized correlation function with
$\gamma(0)=1$, $\gamma(\infty)=0$.
Physically $p_2$ gives the probability that two points a
distance $r$ apart will lie in the first phase.
Two principle fractal types are evident in composites:
surface and volume (or mass) fractals. Each can be characterized by the
behaviour of the normalized correlation function $\gamma(r)$.
Surface fractal behaviour occurs when the interface between the phases
is rough at all scales. In this case Bale and Schmidt~\cite{Bale84} have
shown that $1-\gamma(r) \sim r^{3-D_s}$ as $r\to0$ where
$2\leq D_s < 3$ is the surface-fractal dimension.
Experimental techniques have determined that
a variety of materials exhibit fractal surfaces with $D_s>2$. These
include crushed glass, zeolites~\cite{Avnir83b},
silica gels~\cite{Avnir83b,Schmidt89}, lignite coals~\cite{Bale84}, porous
rocks~\cite{Wong86}, soils~\cite{Bartoli91} and composite
steels~\cite{Su95} (see Korvin~\cite{Korvin92} for a review).
On the other hand, volume (or mass) fractals occur when voids (or inclusions)
are present at all scales. In this case $\gamma(r)\sim r^{-(3-D_v)}$ as
$r\to\infty$ where $D_v < 3$ is the volume-fractal
dimension~\cite{Schaefer84,Wong92}. 
Fractal behavior of this type has been observed in silica
gels~\cite{Schaefer84,Bremer93,Hasmy94} and soils~\cite{Bartoli91}.

\section{Model morphologies: The level cut Gaussian random field}
Random composites with a wide variety of morphological properties 
can be generated by taking level-cuts of a
Gaussian random field (GRF). A simple definition of a GRF is
\begin{equation}\label{defny}
y(\mbox{\boldmath $r$})=\sqrt{\frac{2}{N}}\sum_{i=1}^{N} 
\cos(k_i \hat{\mbox{\boldmath $k$}} \cdot {\mbox{\boldmath $r$}} + \phi_i)
\end{equation} 
where $\phi_i$ is a uniform deviate on $[0,2\pi)$ and
$\hat{\mbox{\boldmath $k$}}_i$ is uniformly distributed on a unit sphere.
The magnitude of the wave vectors $k_i$ are distributed on $[0,\infty)$ with a
probability (spectral) density $P(k)$ ($\int P(k)dk=1$).
A composite material can then be defined by taking the region in space where
$\alpha \leq y(\mbox{\boldmath$r$}) \leq \beta$ as the first phase,
while the two regions contiguous to this $y(\mbox{\boldmath $r$}) < \alpha;\;
y(\mbox{\boldmath $r$}) > \beta$ define a complementary second phase.
In the case $\beta=\infty$ a 1-cut material results; in the case 
$\beta=-\alpha$ a 2-cut material.

The statistical correlation functions of these materials can be
calculated~\cite{Berk87,Berk91,Teubner91}.
The volume fraction is given by
$p=(2\pi)^{-\frac12}\int_\alpha^\beta e^{-\frac12 t^2} dt$ and the two
point correlation function is
\begin{eqnarray}
p_2(r)=&&p^2+\frac{1}{2\pi}\int_0^{g(r)} \frac{dt}{\sqrt{1-t^2}} \times  \left[
\exp\left(-\frac{\alpha^2}{1+t}\right) \right.   \nonumber
\\ && \left.
-\exp\left(-\frac12 \frac{\alpha^2-2\alpha\beta t+\beta^2}{(1-t^2)}\right)
+\exp\left(-\frac{\beta^2}{1+t}\right) \right].
\end{eqnarray}
Here $g(r)=\langle y(0) y(\mbox{\boldmath $r$}) \rangle$ is the 
field-field correlation function and is related to the spectral
density of the field
\begin{equation}
g(r)=\int_0^\infty P(k) \frac{\sin kr}{kr} dk.
\end{equation}
In terms of the normalized 2-point
function $\gamma(r)=(p_2(r)-p^2)/(p-p^2)$ it can be shown that~\cite{Berk91}
\begin{equation} 1-\gamma(r) \sim (1-g(r))^\frac12\;\; r\to 0 \;\;\&\;\;
\gamma(r) \sim g(r) \;\; r\to \infty. \end{equation}
Higher order correlation functions can also be evaluated for level-cut
GRF media~\cite{Roberts95a,Roberts95c,Roberts95d}.
The freedom in specifying $\alpha$, $\beta$ and $P(k)$ (or $g(r)$) allows
a wide variety of materials to be modelled.

\section{Surface fractals}
Recent microstructural studies of composite steels~\cite{Su95} have revealed 
a fractal boundary between a ferrite and pearlite phase.  Using a 
scaling relationship between the area and perimeter of the pearlite grains
it was found that $d_f\approx1.5$. In three-dimensions this implies
a surface fractal dimension of $D_s=1+d_f\approx2.5$. 
Small angle scattering studies have shown that lignite
coal~\cite{Bale84} has $D_s\approx2.5$ and sedimentary sandstones have
$2.25 \leq D_s \leq 2.96$ with many exhibiting $D_s=2.5\pm0.1$~\cite{Wong86}. 
Materials with such a fractal surface can be generated in the
GRF scheme by using, for example, $g(r)=e^{-r/l}$ where
$l$ is a pore/grain length scale which we normalize to unity.
In this case $1-\gamma(r)\sim r^{\frac12}$ as $r\to0$ so that
$D_s=2.5$.  The corresponding spectral density is
\begin{equation} P(k)=\frac{4 k^2}{\pi(1+k^2)^2}. \end{equation}

To account for the fact that physical surfaces only exhibit fractal
scaling down to some finite length scale (e.g.\ the lattice constant)
a cut-off length can be introduced in the model by setting
$P(k)=0$ for $k>K$ (and scaling so $\int P(k) dk=1$).
This leaves the large scale microstructure unchanged and filters
out the high frequency (rougher) sinusoids in Eqn.~(\ref{defny}) above
a length scale $2\pi/K$.  
The magnitude of $K$ is estimated from experimental studies.
In ferritic-pearlite steels~\cite{Su95} the minimum scale reported is
$.1 \mu m $ and the grain size is $O(50\mu m)$ so
$K\approx  2\pi 50/.1=3.1\times 10^3$.
In sandstones~\cite{Wong86} the minimum scale measured is $.5 \times 10^{-3} \mu m$
and the pore-size is around $10 \mu m $ so
$K\approx 2\pi 10/.0005 =1.3\times 10^6$. We assume $K=\infty$
provides a reasonable model of fractal steel and sandstone interfaces.
To clearly demonstrate the
effect of a fractal surface on material properties 
we also consider materials with $K=8$
(providing a smooth interface with $D_s=2$).

\begin{figure}
\begin{center}
\begin{minipage}[hbt]{.60\linewidth}
\begin{minipage}[hbt]{.50\linewidth}
\centering \epsfxsize=.95\linewidth\epsfbox{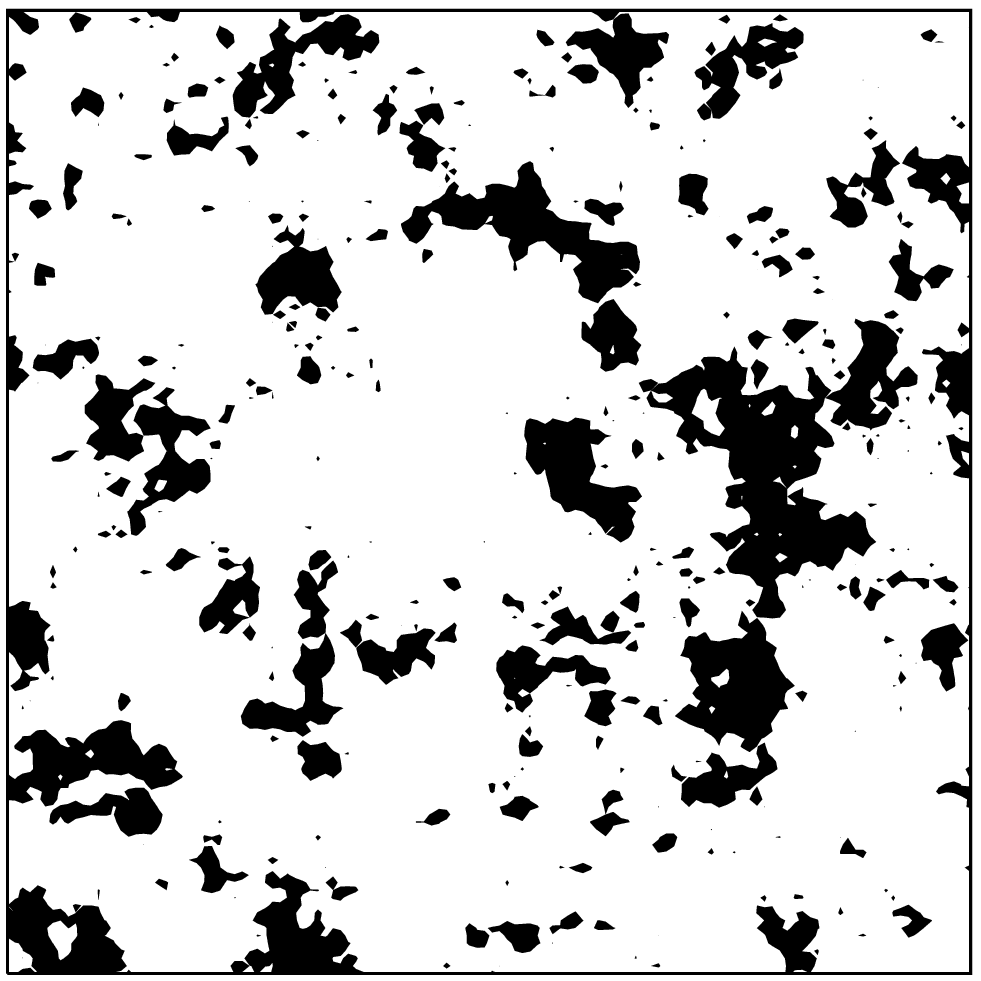}
(a) 1-cut $D_s=2.5$
\end{minipage}\hfill
\begin{minipage}[hbt]{.50\linewidth}
\centering \epsfxsize=.95\linewidth\epsfbox{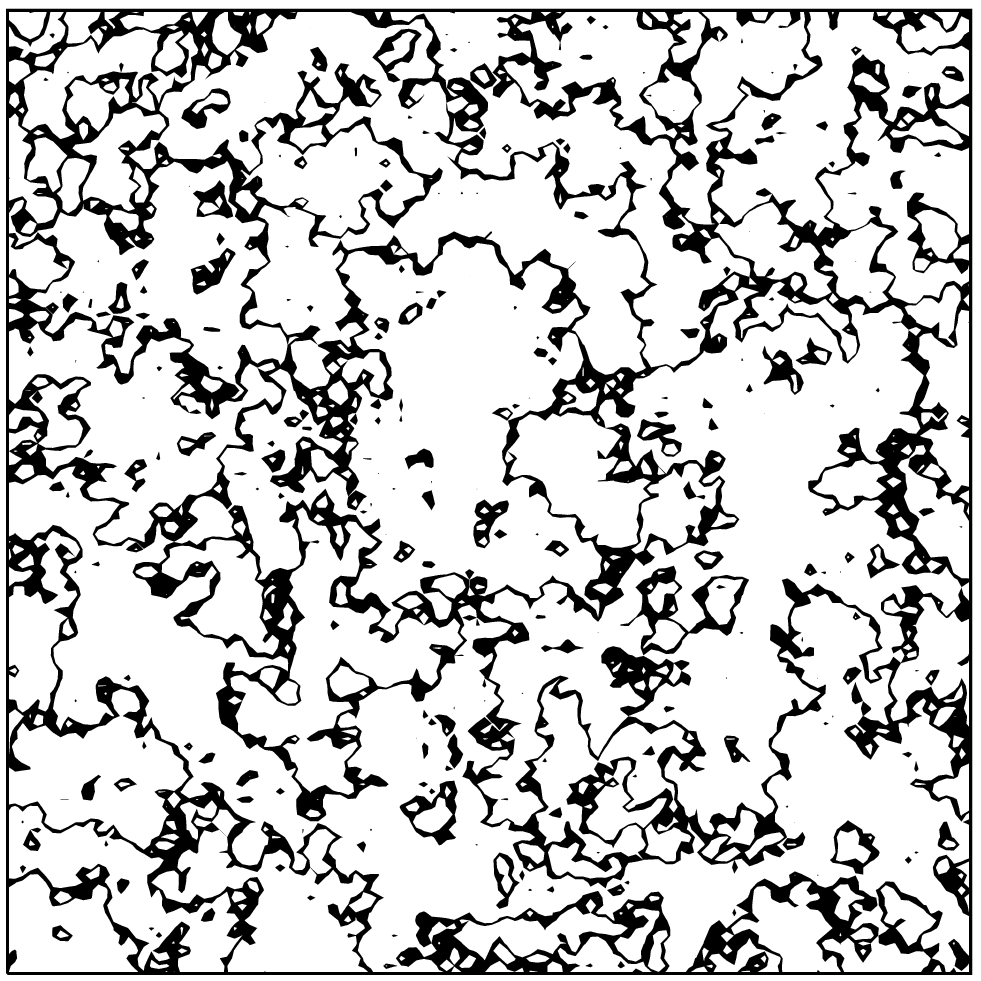}
(b) 2-cut $D_s=2.5$
\end{minipage}\hfill
\begin{minipage}[hbt]{.50\linewidth}
\centering \epsfxsize=.95\linewidth\epsfbox{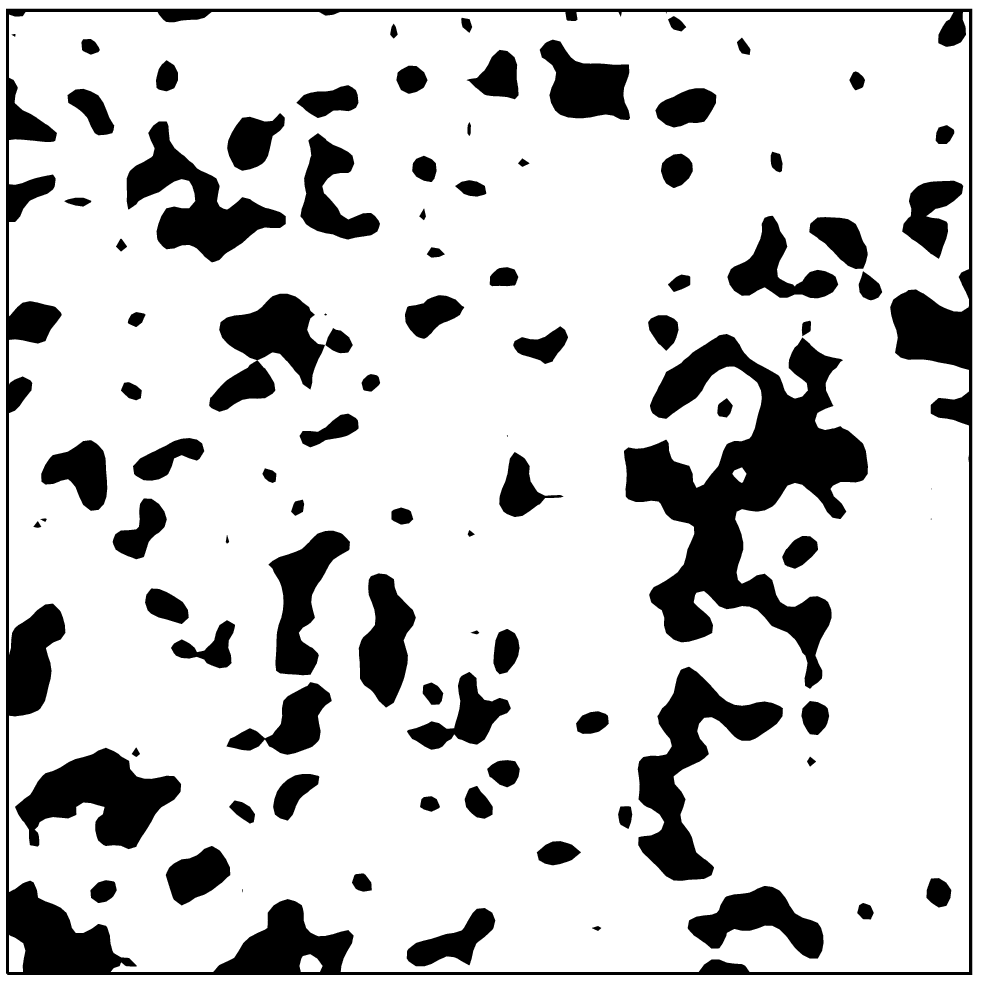}
(c) 1-cut $D_v=1.5$
\end{minipage}\hfill
\begin{minipage}[hbt]{.50\linewidth}
\centering \epsfxsize=.95\linewidth\epsfbox{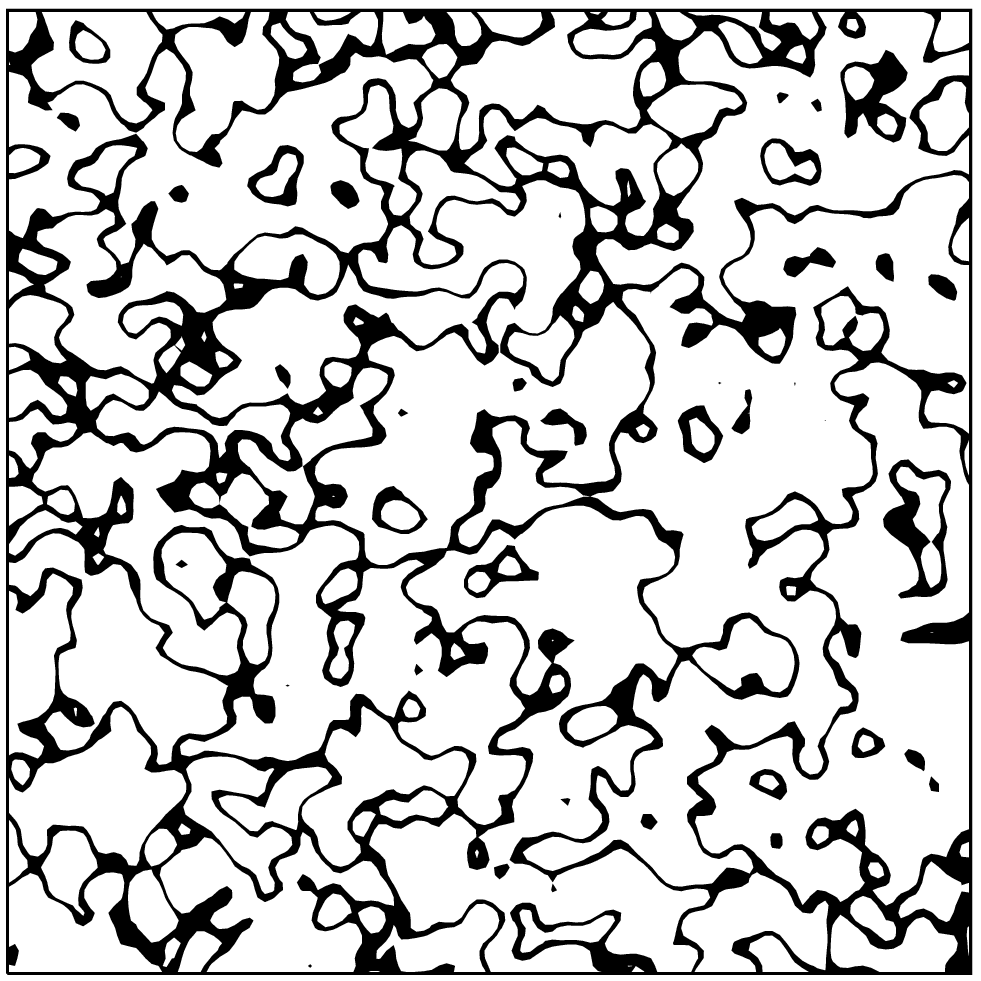}
(d) 2-cut $D_v=1.5$
\end{minipage}\hfill
\end{minipage}
\caption{Cross-sections of the fractal composites generated using
Eqn.~(\protect\ref{defny}). The black region corresponds to Phase 1 at
a volume fraction $p=0.2$. In the 1-cut case $0.84 <y(r)< \infty$ and
in the two cut case $-0.25 <y(r)< 0.25$. \label{cross}}
\end{center}
\end{figure}

\begin{figure}
\begin{minipage}[hbt]{.50\linewidth}
\centering \epsfxsize=.95\linewidth\epsfbox{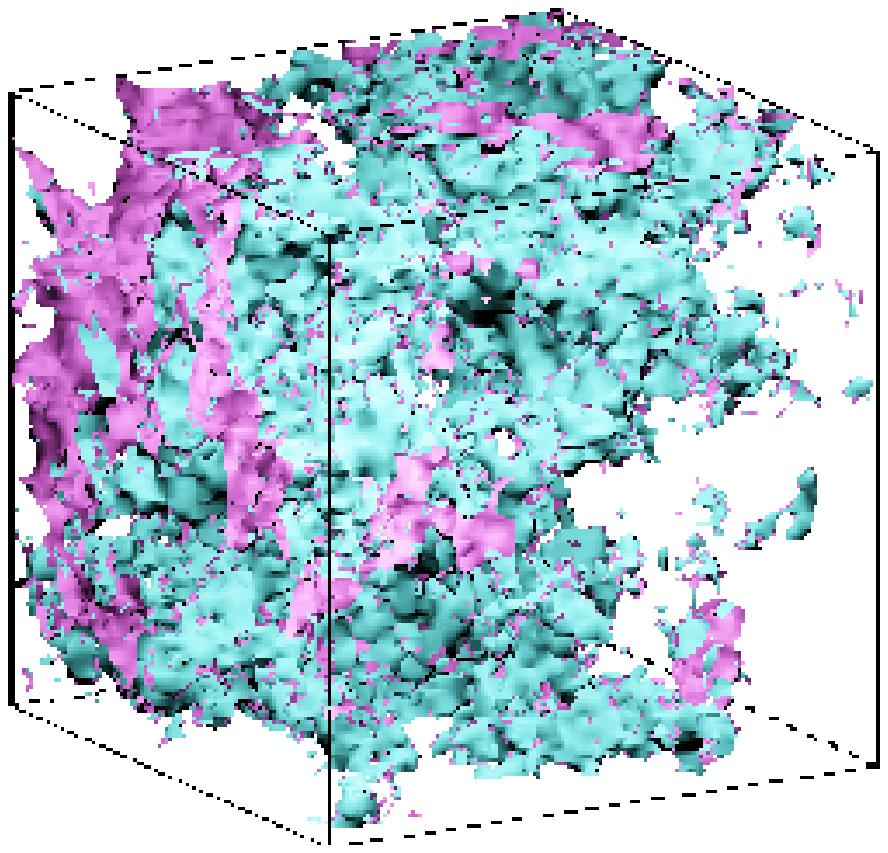}
\end{minipage}\hfill
\begin{minipage}[hbt]{.50\linewidth}
\centering \epsfxsize=.95\linewidth\epsfbox{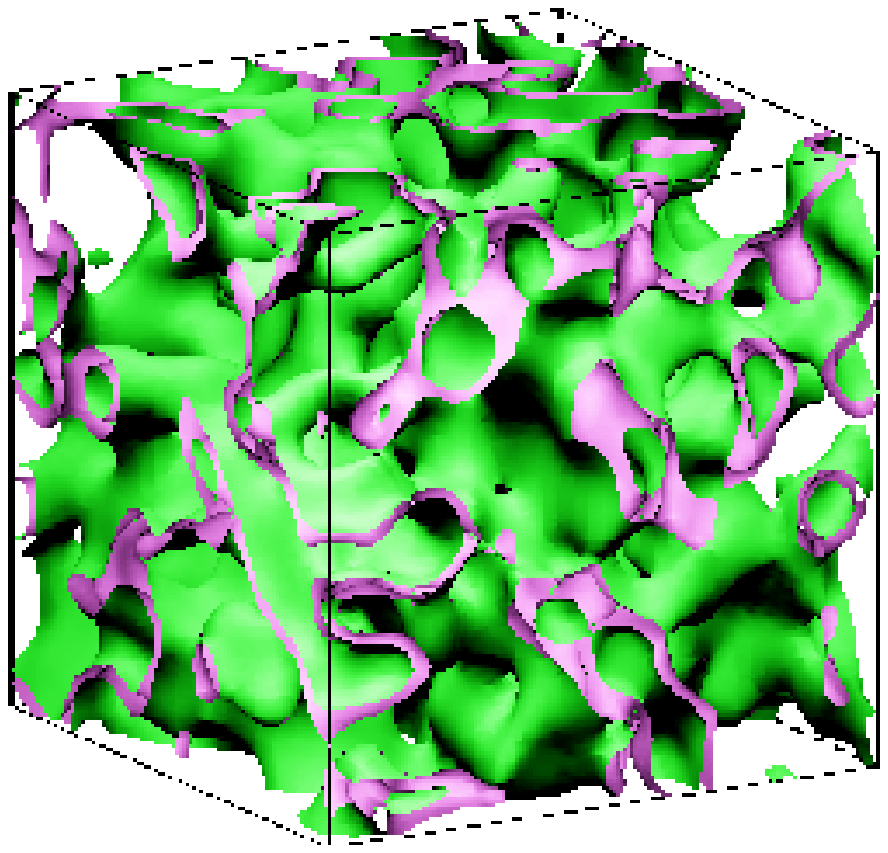}
\end{minipage}\hfill
\caption{The level cut media. (a) A 1-cut surface-fractal
interface ($D_s=2.5$) at $p=0.22$ ($0.75 < y(r) < \infty$); (b)
A 2-cut volume-fractal ($D_v=1.5$) at $p=0.2$ ($-0.25 < y(r) < 0.25$).
\label{3D}}   
\end{figure}

\begin{figure}
\begin{minipage}[hbt]{.50\linewidth}
\centering \epsfxsize=.95\linewidth\epsfbox{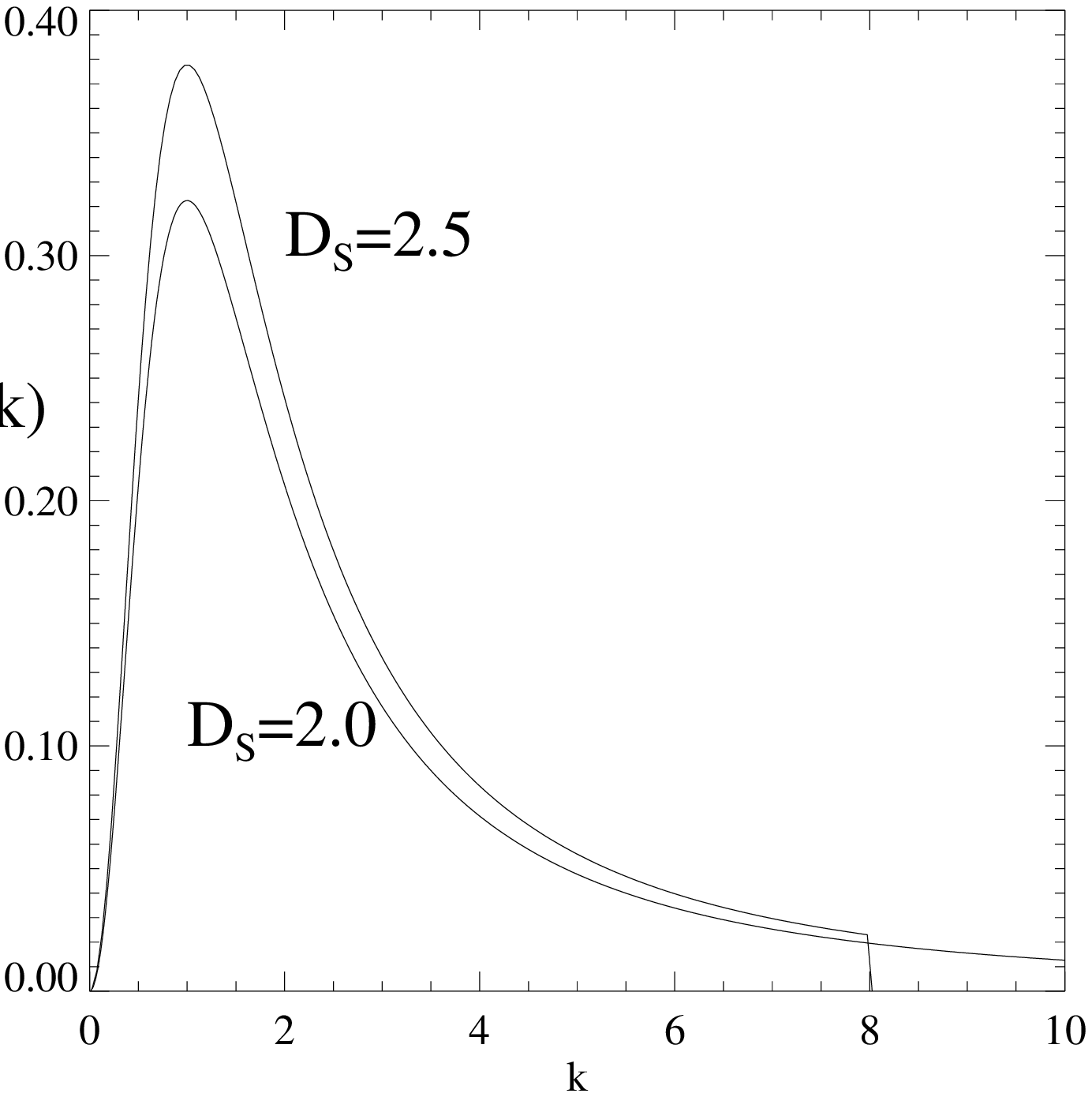}
\end{minipage}\hfill
\begin{minipage}[hbt]{.50\linewidth}
\centering \epsfxsize=.95\linewidth\epsfbox{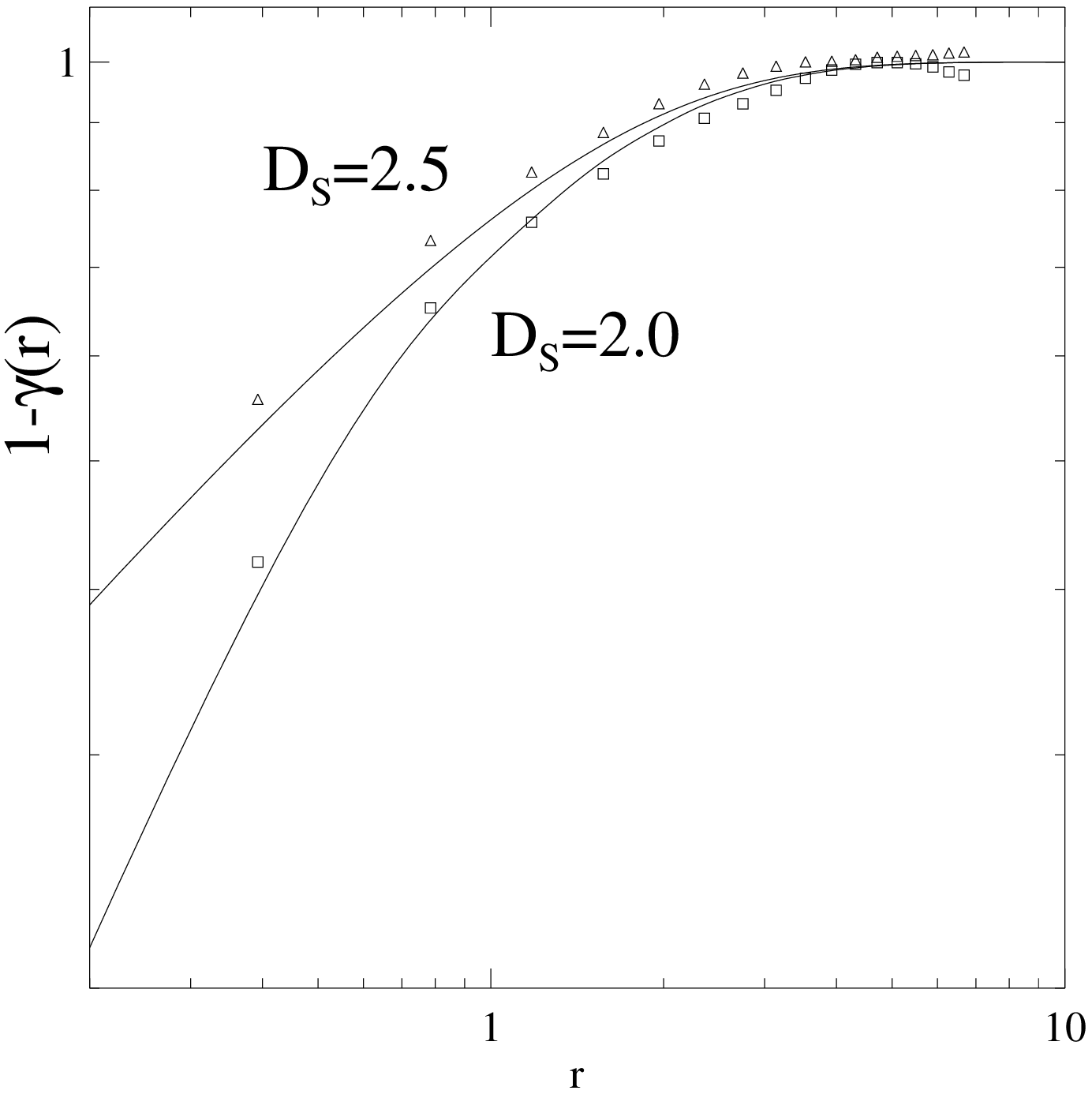}
\end{minipage}\hfill
\caption{The spectral density and normalized 2-point correlation function
of the 1-cut surface fractal media ($p=\frac12$). The symbols represent computational
measurements of $\gamma(r)$ averaged over 10 cross-sections. \label{surgraphs}}
\end{figure}

Cross sections of the surface fractal material are shown in 
Figs.~\ref{cross} (a) \& (b). The morphology of the 2-cut field is
very similar to that observed in steel composites~\cite{Su95}. The roughness
of the interface for the 1-cut case is clearly visible in Fig.~\ref{3D}.
The spectral density and associated 2-point correlation function
of each of the materials are shown in Fig.~\ref{surgraphs}. 
Ten two-dimensional realizations of each model are calculated from
Eqn.~(\ref{defny}) for the case $\alpha=0$ \& $\beta=\infty$ ($p=\frac12$). 
The average value of $\gamma(r)$ is represented by symbols in this
figure: the fractal scaling of $\gamma(r)$ is evident.

\begin{figure}
\centering \epsfxsize=8cm \epsfbox{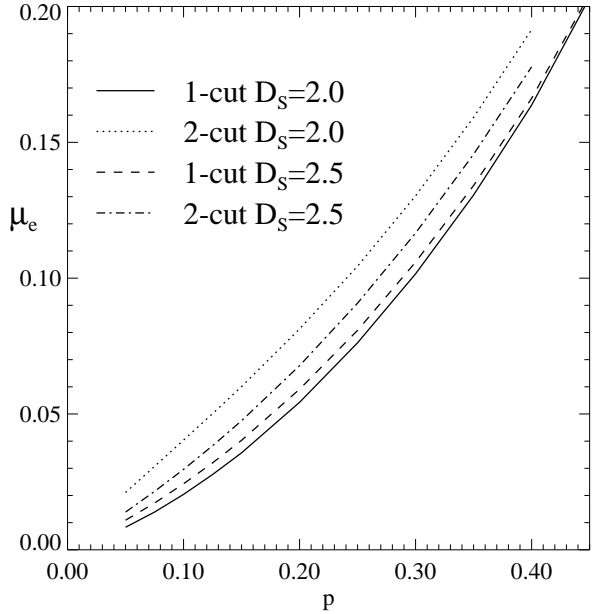}
\caption{Upper bounds~\protect\cite{Milton82b}
on the shear modulus of the surface-fractal media. \label{boundsur}}
\end{figure}

To determine the effect of the fractal microstructure on the
conductivity and elastic moduli of each material we evaluate
the microstructure parameters for both models ($K=\infty,8$).
The results are given in Tables~\ref{tab1cut}~\&~\ref{tab2cut} and bounds
on the shear moduli are shown in Fig.~\ref{boundsur} for the case 
where the the shear and bulk moduli are both equal to 1 in phase 1 and
0 in phase 2. Note that the lower bound vanishes at this contrast.
Curves for the bulk moduli and conductivity are
qualitatively similar. The shear moduli of the fractal and smooth 1-cut
models are quite similar. In contrast the fractal 2-cut model has a
significantly reduced resistance to shear when compared to
the smooth 2-cut model: a rough interface decreases the elastic strength of
a material if the inclusion phase is stronger than the bulk phase.
This result is similar to a trend observed in
composite steels: an increase in $D_s$ leads to a decrease in impact
toughness.


\clearpage

\section{Volume fractals}

Aerogels provide good examples of volume fractal materials with 
$D_v\approx2$~\cite{Schaefer84}. The formation process has been modelled by
the cluster-cluster aggregation model~\cite{Meakin83c,Kolb83,Hasmy94}
which gives $D_v\approx1.7-1.8$.
To see how the level-cut scheme can model volume-fractal materials
consider $\gamma(r)\approx C (r/l)^{-A}$ for $r\to\infty$.
The volume of solid within a radius $R$ is~\cite{Schaefer84}
\begin{eqnarray} \nonumber
V(R)&=&\int_0^R dv p_2(r)/p=\int_0^R dr 4\pi r^2\left((1-p)\gamma(r)+p \right)
\\ &\approx& \frac{(1-p)4\pi C l^A}{3-A} R^{3-A} +p\frac{4\pi}{3} R^3 .
\label{volscale}
\end{eqnarray}
Now the the former term dominates for  
$R < R_s = \left( \frac{3 (1-p) C}{p(3-A)} \right)^{1/A}l\sim p^{-1/A}l$
where $R_s$ is a saturation scale and $V(R) \sim R^{3-A} = R^{D_v}$. Thus the
volume fractal dimension is $D_v=3-A$.
A simple field-field correlation function which gives rise to this behaviour in
the level-cut GRF is
\begin{equation} 
g(r)=\left(1+\frac2A\frac{r^2}{l^2}\right)^{-A/2}. \label{volfrac}
\end{equation} 
This field-field correlation function actually leads to a
spectral density
\begin{equation}
P(k)=\frac1{(2\pi)^3}\int_0^\infty 4\pi r^2 g(r) \frac{\sin kr}{kr} dr
\end{equation}   
which is not strictly positive. 
Therefore we define a new $P(k)$ which is set to zero
for all $k$ beyond the first point $K$ at which $P(k)<0$ (and re-scale
to ensure $\int P(k) dk =1$). This modification does not change the
behaviour of the model at large $r$.
In aerogels $l$ (which we normalize to unity) is
related to the length scale of the monomers. In experiments the saturation
scale is $O(10$nm)~\cite{Schaefer84} and in a
recent model of a silica gel based on the cluster-cluster aggregation
algorithm the saturation scale is around six times the particle
diameter at concentrations of $c=0.05$~\cite{Hasmy94}. This is consistent
with Eqn.~(\ref{volscale}): $R_s \sim p^{-1/A}=0.05^{-1/1.7}=5.8$.
Although the level-cut materials generated using Eqn.~(\ref{volfrac})
has fractal scaling over the correct range of $r$, it does not necessarily
follow that this is an appropriate model for aerogels. It does, however,
allow the dependence of properties on $D_v$ to be gauged within the 
level-cut scheme.

\begin{table}
\caption{\sl The microstructure parameters the fractal and
regular composites for 1-level cut GRF's ($\beta=\infty$).}
\label{tab1cut}
\begin{center}
\begin{tabular}{|c|c|c|c|c|c|c|c|c|}
\hline
\multicolumn{1}{|c|}{} &
\multicolumn{8}{|c|}{Model} \\
\hline
\multicolumn{1}{|c|}{}   &
\multicolumn{2}{|c|}{$D_s=2.5$}   &
\multicolumn{2}{|c|}{$D_s=2.0$}   &
\multicolumn{2}{|c|}{$D_v=1.5$}   &
\multicolumn{2}{|c|}{$D_v=3.0$}   \\
\hline
\multicolumn{1}{|c|}{p} &
\multicolumn{1}{|c|}{$\zeta_1$} &\multicolumn{1}{|c|}{$\eta_1$} &
\multicolumn{1}{|c|}{$\zeta_1$} &\multicolumn{1}{|c|}{$\eta_1$} &
\multicolumn{1}{|c|}{$\zeta_1$} &\multicolumn{1}{|c|}{$\eta_1$} &
\multicolumn{1}{|c|}{$\zeta_1$} &\multicolumn{1}{|c|}{$\eta_1$} \\
\hline
0.05 &0.294 & 0.247 & 0.215 & 0.165 & 0.220 & 0.172 & 0.197 & 0.148 \\
0.10 &0.319 & 0.276 & 0.258 & 0.212 & 0.263 & 0.219 & 0.243 & 0.197 \\
0.15 &0.342 & 0.305 & 0.294 & 0.253 & 0.299 & 0.260 & 0.281 & 0.239 \\
0.20 &0.366 & 0.332 & 0.327 & 0.291 & 0.330 & 0.296 & 0.316 & 0.279 \\
0.25 &0.388 & 0.360 & 0.358 & 0.327 & 0.360 & 0.331 & 0.348 & 0.317 \\
0.30 &0.411 & 0.390 & 0.387 & 0.363 & 0.388 & 0.365 & 0.380 & 0.354 \\
0.35 &0.433 & 0.415 & 0.416 & 0.396 & 0.417 & 0.398 & 0.411 & 0.391 \\
0.40 &0.456 & 0.442 & 0.444 & 0.431 & 0.444 & 0.431 & 0.441 & 0.427 \\
0.50 &0.500 & 0.500 & 0.500 & 0.500 & 0.498 & 0.496 & 0.500 & 0.500 \\
\hline
\end{tabular}
\end{center}
\end{table}
\begin{table}
\caption{\sl The microstructure parameters the fractal and
regular composites for 2-level cut GRF's ($\beta=-\alpha$).}
\label{tab2cut}
\begin{center}
\begin{tabular}{|c|c|c|c|c|c|c|c|c|}
\hline
\multicolumn{1}{|c|}{} &
\multicolumn{8}{|c|}{Model} \\
\hline
\multicolumn{1}{|c|}{}   &
\multicolumn{2}{|c|}{$D_s=2.5$}   &
\multicolumn{2}{|c|}{$D_s=2.0$}   &
\multicolumn{2}{|c|}{$D_v=1.5$}   &
\multicolumn{2}{|c|}{$D_v=3.0$}   \\
\hline
\multicolumn{1}{|c|}{p} &
\multicolumn{1}{|c|}{$\zeta_1$} &\multicolumn{1}{|c|}{$\eta_1$} &
\multicolumn{1}{|c|}{$\zeta_1$} &\multicolumn{1}{|c|}{$\eta_1$} &
\multicolumn{1}{|c|}{$\zeta_1$} &\multicolumn{1}{|c|}{$\eta_1$} &
\multicolumn{1}{|c|}{$\zeta_1$} &\multicolumn{1}{|c|}{$\eta_1$} \\
\hline
0.05 &0.402 & 0.351 & 0.786 & 0.613 & 0.768 & 0.589 & 0.802 & 0.627 \\
0.10 &0.415 & 0.369 & 0.706 & 0.516 & 0.689 & 0.500 & 0.727 & 0.535 \\
0.15 &0.431 & 0.387 & 0.656 & 0.480 & 0.636 & 0.450 & 0.684 & 0.501 \\
0.20 &0.449 & 0.402 & 0.628 & 0.473 & 0.615 & 0.470 & 0.657 & 0.491 \\
0.25 &0.463 & 0.430 & 0.612 & 0.478 & 0.602 & 0.473 & 0.642 & 0.495 \\
0.30 &0.481 & 0.451 & 0.605 & 0.491 & 0.596 & 0.488 & 0.636 & 0.508 \\
0.35 &0.497 & 0.474 & 0.603 & 0.510 & 0.596 & 0.506 & 0.634 & 0.526 \\
0.40 &0.515 & 0.495 & 0.607 & 0.533 & 0.601 & 0.529 & 0.636 & 0.546 \\
\hline
\end{tabular}
\end{center}
\end{table}

\begin{figure}
\begin{minipage}[hbt]{.50\linewidth}
\centering \epsfxsize=.98\linewidth \epsfbox{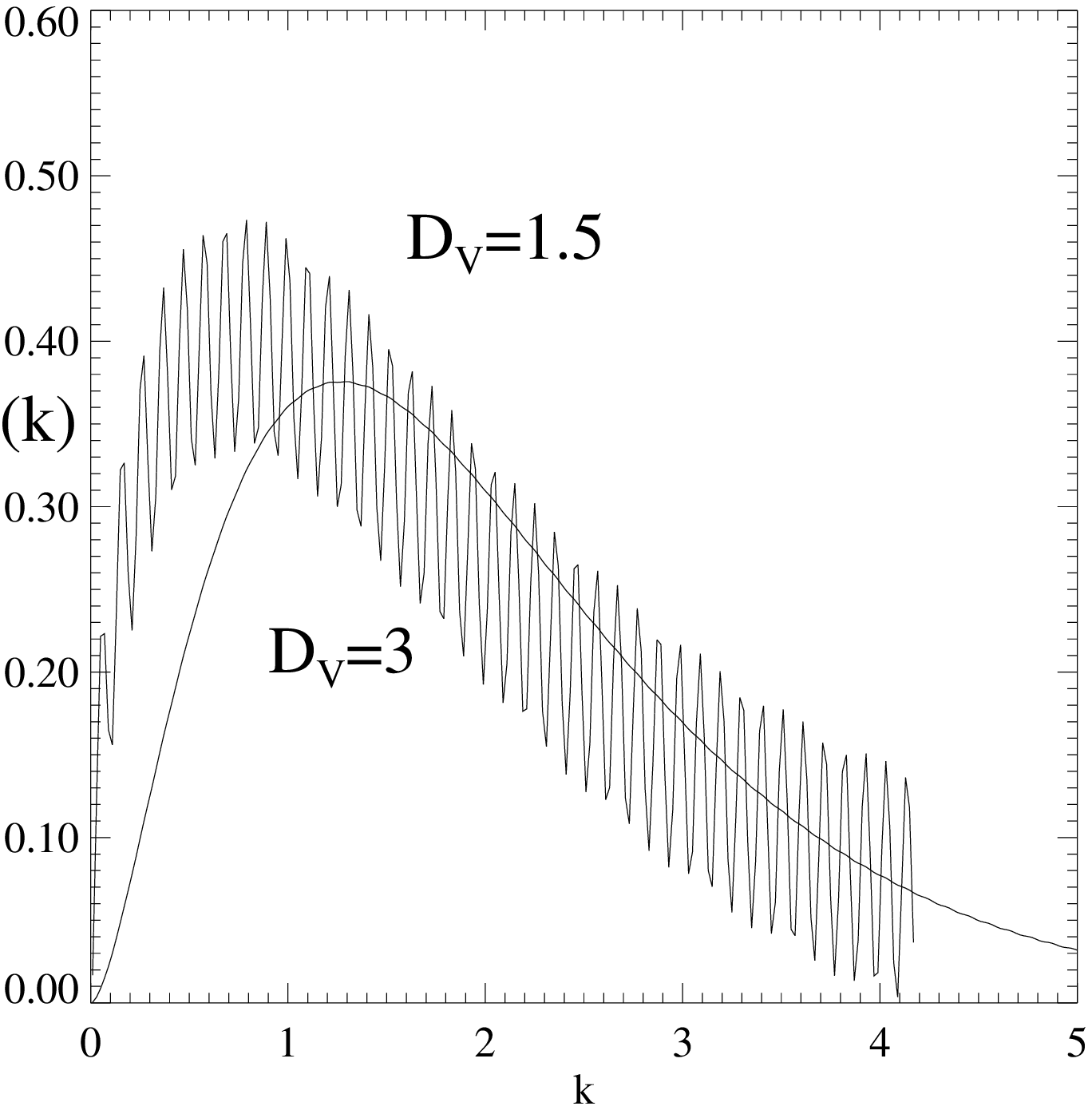}
\end{minipage}\hfill
\begin{minipage}[hbt]{.50\linewidth}
\centering \epsfxsize=.98\linewidth \epsfbox{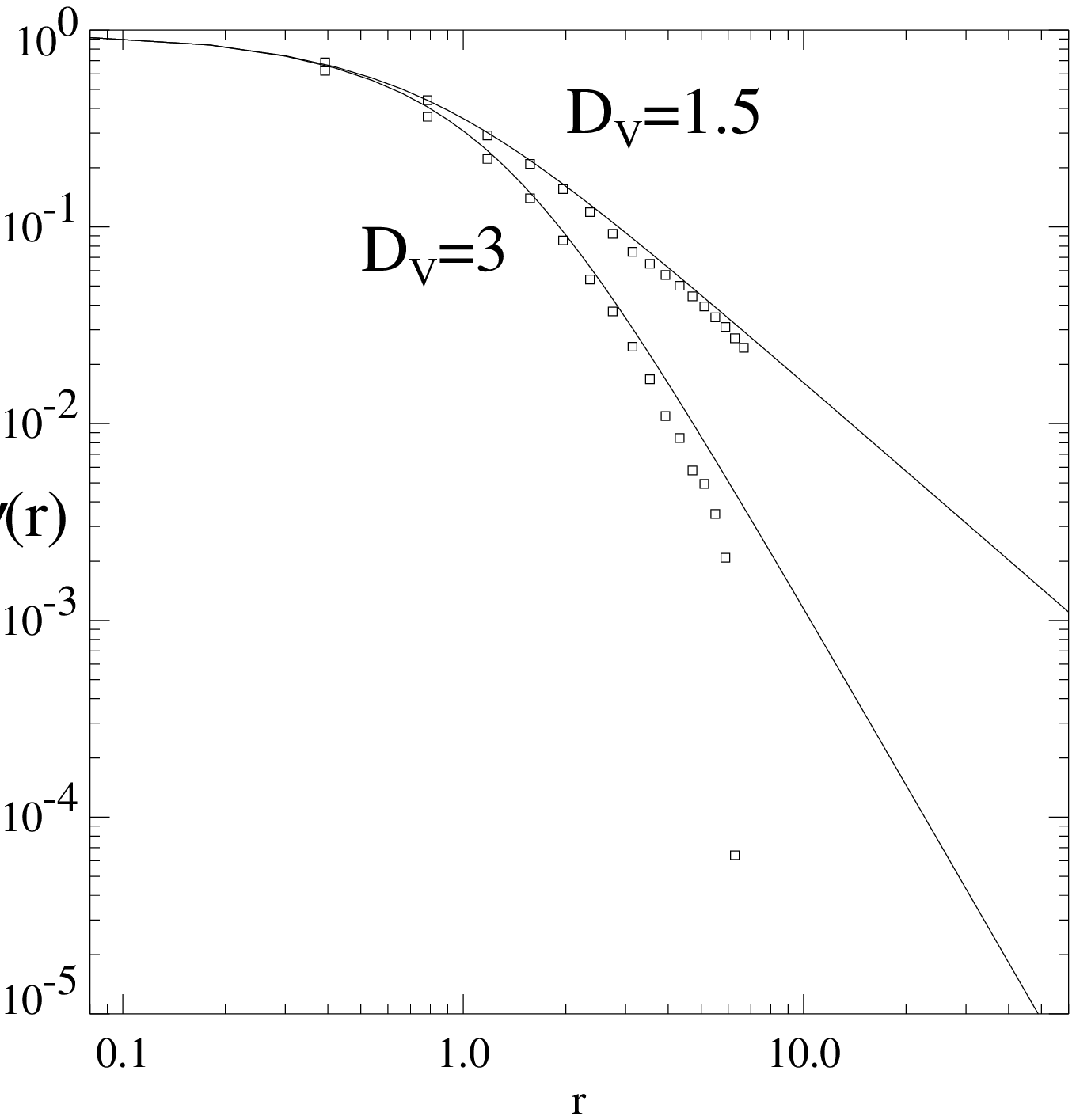}
\end{minipage}\hfill
\caption{The spectral density and normalized 2-point correlation function
of the 1-cut volume fractal media ($p=\frac12$). The symbols represent computational
measurements of $\gamma(r)$ averaged over 50 cross-sections. \label{volgraphs}}
\end{figure}

We choose a fractal $A=1.5$ ($D_v=1.5$) and non-fractal case $A=3$
(so the second term in Eqn.~(\ref{volscale})  dominates and $D_v=3$) to
determine the effect of volume fractal behaviour on material properties.
The functions $P(k)$ are shown in Fig.~\ref{volgraphs}. 
We have generated realizations of the materials using
Eqn.~(\ref{defny}). Cross-sections of the
model for the case $D_v=1.5$ are shown for the 1-cut and 2-cut cases in
Figs.~\ref{cross} (c) \& (d). A three dimensional representation of the
2-cut interface is shown in Fig.~\ref{3D}.
The 2-point function of each model has been measured
for the case $p=\frac12$ and averaged over 50 cross-section realizations
of each model.  The results, plotted in Fig.~\ref{volgraphs}, show that
the level-cut GRF model has volume-fractal scaling.

\begin{figure}
\centering \epsfxsize=8cm \epsfbox{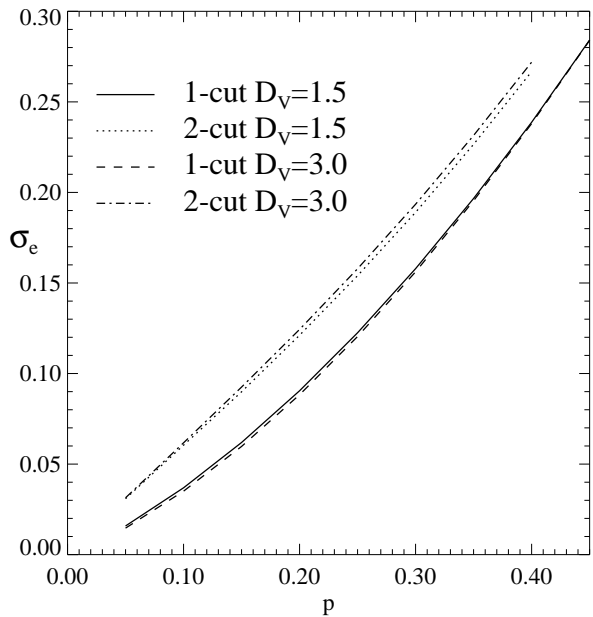}
\caption{Upper bounds~\protect\cite{Beran65a,Milton81a}
on the conductivity of the volume-fractal media. \label{boundvol}}
\end{figure}

The microstructure parameters for both $D_v=1.5$ and $D_v=3.0$ are given
in Tables~\ref{tab1cut}~\&~\ref{tab2cut} and bounds
on the conductivity are shown in Fig.~\ref{boundvol} for the case 
$\sigma_{1,2}=1,0$ (the lower bounds vanish at this contrast).
As we would expect the sheet-like
nature of the structures in the 2-cut media are much better conductors than
the node/bond-like structures present in the 1-cut media.
We find virtually no difference in the fractal and non-fractal materials
in both the 1-cut and 2-cut cases. This surprising result suggests that
the properties of volume-fractal composites (such as aerogels) 
are not explicitly dependent on the fractal dimension. 


\section{Conclusion}

We investigate the influence of fractal structure on material
properties.
We have calculated rigorous bounds on the conductivity and elasticity of
fractal media generated using the level-cut random field model. The
behaviour of the bounds indicates that a fractal interface plays a minimal
role in the properties of 1-cut media. For the two-cut model, which mimics
the microstructure of both foams~\cite{Roberts95b} and porous
rocks~\cite{Roberts95d} a much stronger influence is observed. 
In contrast, varying the volume-fractal dimension
of both 1-cut and 2-cut media has little effect on the property bounds.
The latter result indicates that the remarkable properties~\cite{Lu92} of
aerogels are influenced more by the fact that they
contain very well connected structures at high porosities, rather than their
fractal characteristics~\cite{Schaefer84,Hasmy94}.
In the future we shall utilise a range of microstructural
studies~\cite{Hasmy94} to develop a more appropriate model of aerogel
structure.
This will allow a more rigorous
comparison between model and experimental properties.


\end{document}